\documentclass[reprint,3p,12pt]{elsarticle}
\usepackage{amsmath,amssymb,graphicx}

\newtheorem{Algorithm}{Algorithm}
\newtheorem{thm}{Theorem}
\newtheorem{defn}[thm]{Definition}

\begin{document}

\begin{frontmatter}
\title{Computation of the shortest path between two curves on a  parametric surface by  geodesic-like method}

\author[Chen1]{Wen-Haw Chen}
\ead{whchen@thu.edu.tw}
\author[Chen2]{Sheng-Gwo Chen\corref{cor}}\ead{csg@mail.ncyu.edu.tw}

 \cortext[cor]{Corresponding author: S.-G. Chen}
\address[Chen1]{Department of Mathematics, Tunghai University, Taichung 40704, Taiwan.}

\address[Chen2]{Department of Applied Mathematics, National Chiayi University,  Chia-Yi 600,
Taiwan.}

\begin{abstract}
In this paper, we present the geodesic-like algorithm for the
computation of the shortest path between two objects on NURBS
surfaces and periodic surfaces. This method can improve the distance
problem not only on surfaces but in $\mathbb{R}^3$. Moreover, the
geodesic-like algorithm also provides an efficient approach to
simulate the minimal geodesic between two holes on a NURBS surfaces.

\end{abstract}

\begin{keyword}
Distance \sep Geodesic-like curves \sep Orthogonal projection \sep
Parametric surface \sep Shortest path
\end{keyword}

\end{frontmatter}

\section{Introduction}

Computing the distance between two objects on a surface plays an
important role in many fields such as   CAD, CAGD, robotics and
computer graphics etc. In the Euclidean 3-space $\mathbb{R}^3$, it
has a simple mathematical presentation as following:
\begin{equation}
 \min_{p \in c_1,\; q \in c_2}  \|p-q\|,
\end{equation}
where $c_1$ and $c_2$ are  two objects in $\mathbb{R}^3$.

Although this representation is simple, however, it is hard to
improve in general. The simplest case is the distance between two
points and it can be estimated exactly by the Pythagorean theorem.
If  only one object is a point, then this problem is equivalent to
the orthogonal projection problem, which has many
applications \cite{Piegl2, Penga}. Many investigators have
investigated the orthogonal projection problem.
Chen et al. \cite{Chen2,Chen3}, MaYL et al. \cite{MaYL} and
Selimovic et al. \cite{Selimovic} presented some effective methods that
improve the distance problem between a point and a NURBS curve.
Hu et al. \cite{Hu} developed a good method to improve the orthogonal
projection onto curves and surfaces. For the case that none of these
objects is a single point, Kim \cite{Kim} presented a method to
estimate the distance between a canal surface and a simple surface
in 2003, while Chen et al. \cite{Chen4} improved this problem on two
implicit algebraic surfaces in 2006.

Unlike one can find many methods to investigate the distance problem
in $\mathbb{R}^3$, there are few methods to study that on a curved
surface. Maekawa \cite{Maekawa} presented a very good method for
solving the shortest path and the orthogonal projection problems on
free-form parametric surfaces. Generally, the distance problem on a
regular surface is more complicated than that in $\mathbb{R}^3$,
even though the distance is just between two points on the surface,
which is equivalent to find the length of the shortest path between
them. We can find more information in reference \cite{Patrikalakis}.
This classical problem has many applications, such as in object
segmentation, multi-scale image analysis and CAD etc. \cite{Caselles,
Kimmel, Sanchez}. There are also many  methods to estimate the
shortest path on triangular mesh \cite{Martinez, Surazhsky},
polyhedral \cite{Kanai, Polthier} and regular surface \cite{Kasap,
Ravi3} etc. These methods can be extend to improve the distance
between one point and one curve or between two curves on surface but
they are not effective methods.

In 2009, Chen \cite{Chen5} presents a new method to find geodesics
on surfaces by the system of geodesic-like equations
\begin{equation} \nabla E(u_i,v_j) = 0.
\end{equation}
In this paper, the distance problem on NURBS surfaces and parametric
surfaces will be improved by the geodesic-like method with B-spline
basis. In fact, the geodesic-like method can also estimate the
distance between two objects in $\mathbb{R}^3$ but its efficiency is less than
other algorithms that we known.

This paper is structured as follows. Section 2 describes the
definition of the distance problem on regular surfaces and the
notion of geodesic-like curves. We shall present our geodesic-like
algorithm to estimate the distance between two objects on surfaces, especially on periodic surfaces, in section 3.
Section 4 presents some examples about the distance problem on NURBS surfaces by simulations. Finally, we illustrate a discussion about our method and conclude this paper in Section 5.

\section{Preliminaries}

Let us introduce the distance problem between two curves on a
regular surface and the system of geodesic-like equations in this
section. Suppose that $S$ is a regular surface and $(U,\mathbf{x})$
is a system of coordinates on $S$. A curve
$\gamma(t)=(x_1(t),x_2(t))$ is a geodesic curve in $(U,\mathbf{x})$
on $S$ if it satisfies the system of geodesic equations \cite{Carmo}
\begin{equation}\label{geodesic_equation}
\frac{d^2x_k}{dt^2} + \sum_{i,j}\Gamma_{ij}^k
\frac{dx_i}{dt}\frac{dx_j}{dt} = 0, \mbox{~~~~} k=1,2.
\end{equation}

From calculus of variation, a geodesic has an equivalent definition as below.

\begin{defn}
 A geodesic on a regular surface is a critical point of the energy variations. That is, the geodesic $\gamma(t)$ is a
critical point of the energy function
\begin{equation}\label{variation}
E(s) = \frac{1}{2}\int_{a}^{b} \| \frac{\partial f}{\partial t}(s,t)
\|^2 dt,
\end{equation}
where $f(s,t)$ is any proper variation of $\gamma(t)$.
\end{defn}

Here is a basic relationship between the length and energy functions.
\begin{thm}\label{geodesic_thm}
Let $S$ be a regular surface and $p,q \in S$ be two distinct points. If $\alpha$ is a shortest path between $p$ and $q$ on $S$, then
$\alpha$ is a geodesic on $S$ which pass through $p$ and $q$. That is, the geodesic $\gamma(t)$ is a critical point of the length function
\begin{equation}
L(s) = \int_a^b \| \frac{\partial f}{\partial t}(s,t) \| dt.
\end{equation}
\end{thm}

The distance between two points on a surface $S$ is defined by the length of minimum path on $S$ from $p$ to $q$. Then
\begin{equation}
 d(p,q) = \min _{ \gamma \in \Gamma} L(\gamma),
\end{equation}
where $\Gamma$ is the set of all paths on $S$ from $p$ to $q$ and $L(\gamma)$ is the length of the curve $\gamma$ on $S$. From Theorem \ref{geodesic_thm}, the set $\Gamma$ can be only considered the set of all geodesic on $S$ from $p$ to $q$.

Consider the parametric surface $S$ with a parametrization
$\mathbf{x}: U  \rightarrow \mathbb{R}^3$ and two curves
$\mathbf{c}_1$ and $\mathbf{c}_2$ on $S$. For simplicity, we denote
$c_1, c_2 : [a,b] \rightarrow U$ such that
$\mathbf{c}_1=\mathbf{x}(c_1([a,b]))$ and
$\mathbf{c}_2=\mathbf{x}(c_2[a,b]))$. Thus the distance between
$\mathbf{c}_1$ and $\mathbf{c}_2$ on $S$ can be computed by
\begin{equation}\label{distance1}
d(\mathbf{c}_1,\mathbf{c}_2) = \min_{s,t \in [a,b]}  d(\mathbf{x}(c_1(s)),\mathbf{x}(c_2(t)) ).
\end{equation}

That is, $d(\mathbf{c}_1,\mathbf{c}_2)$ is the length of minimal
geodesic from $\mathbf{c}_1$ to $\mathbf{c}_2$. Equation
(\ref{distance1}) introduces  a simple algorithm to improve this
distance problem but it is too expansive. Let us describe it
roughly.

\begin{Algorithm}\label{simplestalgorithm}
First, we digitize the curves $\mathbf{c}_1$ and $\mathbf{c}_2$ to
two sequences of points, $\{p_i\}_{i=0}^m$ and $\{q_j\}_{j=0}^n$,
respectively. For each $i,j$, estimating the minimal geodesic
$\gamma_{ij}$ between $p_i$ and $q_j$. Then the shortest path in $\{
\gamma_{ij} \}_{(i,j)=(0,0)}^{(m,n)}$ approaches the minimal
geodesic between $\mathbf{c}_1$ and $\mathbf{c}_2$ on surface $S$
when $m,n$ are large enough. Of course its length approaches the
minimum distance between $\mathbf{c}_1$ and $\mathbf{c}_2$ on $S$.

\end{Algorithm}

Solve the geodesic between two fixed points is crucial to solve
Algorithm \ref{simplestalgorithm}. One can find many effective
methods in the references \cite{Hotz, Kasap, Martinez, Polthier,
Ravi3, Surazhsky}. However, if the numbers of $\{p_i\}$ and
$\{q_j\}$ are large, this algorithm becomes very slow. In fact,
Algorithm \ref{simplestalgorithm} is the simplest and the slowest
method to improve this problem.\\

We will improve the geodesic problem by the notion of geodesic-like
curves \cite{Chen5}.

\begin{defn}\label{BGeodesic}
 Let $\mathbf{x}(u,v)$ be a parametrization of a regular surface
$S$, $\mathbf{x}:U \subset \mathbb{R}^2 \rightarrow S$. A curve
$\tilde{\alpha}(s)$ on $U$ is called a geodesic-like curve of order $n+1$ on $S$ if
$\tilde{\alpha}(s) = \sum_{i=0}^n N^n_i(s)(\tilde{u}_i,\tilde{v}_i)$ is a
B-spline curve and satisfies the system of geodesic equations
\begin{equation}\label{standard_geodesic_like} (\nabla E
)(\tilde{u}_i,\tilde{v}_j) = 0, \end{equation} where

$$ E (u_i,v_j) = \frac{1}{2} \int _{a}^{b} \|\mathbf{x}(\alpha(t)) \|^2 dt $$
is the energy function of curve
$$ \alpha(t) = \sum_{i=0}^n N_i^n(t)( u_i,v_i) $$
and $(\nabla E )(u_i,v_j) $  is the gradient of $E(u_i,v_j)$.
\end{defn}

Equation (\ref{standard_geodesic_like}) is called the system of
standard geodesic-like equations. Although the system of
geodesic-like equations are integral equations, they can be improved
by the Newton's method, the iterator method or other numerical
methods \cite{Gutierrez, Kasap, Polak1, Press, Ye} effectively.

Since any piecewise differential curve can be approximated by the
B-spline curves, a geodesic-like curve approaches a geodesic on $S$
when the order of the geodesic-like curve is large enough. In the
other words, we can estimate the distance between two points on $S$
via the minimal geodesic-like curves. We summarize this property as
follows.

\begin{thm}\label{geodesic_like_approach_geodesic}
Let $S$ be a parametric surface and let $\gamma:[0,1]\to S$ be a
geodesic. Assume that the curve $\alpha_n = \sum_{i=0}^n
N^n_i(t)(u_i,v_i)$ is the geodesic-like curve between $\gamma (0)$
and $\gamma(1)$ for each positive integer $n \geq 2$. Then
\begin{equation}
\lim_{n \rightarrow \infty} \alpha_n = \gamma.
\end{equation}
\end{thm}

\section{Distance problem by geodesic-like algorithm}

The system of geodesic-like equations provides an elegant method to
improve the distance problem between two objects on surfaces. We are
now in a position to introduce this method in this section. The
parametrization $\mathbf{x}$ on $S$ is defined on $U = [a,b] \times
[c,d]$. That is
$$ \mathbf{x}: [a,b]\times [c,d] \rightarrow S \subset \mathbb{R}^3.$$
 Let  $c_1$ and $c_2$ be two differentiable parameterized curves on $S$ and
$$ \begin{array}{l}
c_1(s) : [0,1] \rightarrow [a,b]\times[c,d] \cr

c_2(t) : [0,1] \rightarrow [a,b]\times[c,d].
\end{array}$$
Thus $\mathbf{c}_1=\mathbf{x}(c_1([0,1]))$ and $\mathbf{c}_2=\mathbf{x}(c_2([0,1]))$
are two curves on $S$. To exclude the zero distance case from our
consideration, we can assume that the two curves have no
intersection. Denote $c_1(s)=(u_0(s),v_0(s))$ and
$c_2(t)=(u_n(t),v_n(t))$ where $u_0$, $u_n:[0,1]\rightarrow [a,b]$
and $v_0$, $v_n:[0,1]\rightarrow [c,d]$ are all differentiable
functions. Note that a B-spline curve $\alpha$ from $[0,1]$ into $[a,b]
\times [c,d]$ with $\alpha(0) \in c_1$ and $\alpha(1) \in c_2$ always has the
form as
\begin{equation}\label{parameter_c}
\begin{array}{ll}
\alpha(x) & = \sum_{i=1}^{n-1}N^n_i(x)(u_i,v_i) + N^n_0(x)c_1(s) +
N^n_n(x)c_2(t) \cr
 & = \sum_{i=1}^{n-1}N^n_i(x)(u_i,v_i) + N^n_0(x)(u_0(s),v_0(s)) +
N^n_n(x)(u_n(t),v_n(t)) \end{array}
\end{equation}
where $x \in [0,1]$.

Hence, we rewrite the system of geodesic-like equations to the
following three different forms. These formulas improve the distance
between two curves on $S$, the orthogonal projection problem on $S$
and the shortest path between two points on $S$, respectively.

\begin{description}
\item[The system of geodesic-like equations between two curves:]
From the equation (\ref{parameter_c}), the parameters of the energy
function $E$ are $s,t,u_1, \cdots, u_{n-1}, v_1, \cdots, v_{n-1}$.
The system of geodesic-like equations between two curves can be
rewritten as

\begin{equation}\label{geodesic_like_two_curves}
(\nabla E) = (
E_{s},E_{t},E_{u_1},E_{u_2},\cdots,E_{u_n-1},E_{v_1},E_{v_2},\cdots,E_{v_n-1}
) = 0
\end{equation}

\item[The system of geodesic-like equations between one point and one curve:]
If $\mathbf{c}_1$ is a constat curve on $S$, then the derivative of $E$ about $t$ is vanish. Thus we obtain the geodesic-like equation between one point and one curve.
\begin{equation}\label{geodesic_like_point_curve}
( \nabla E) = (
E_{t},E_{u_1},E_{u_2},\cdots,E_{u_n-1},E_{v_1},E_{v_2},\cdots,E_{v_n-1}
) = 0.
\end{equation}
Of course, The orthogonal projection projection problem on surface
cab be improve by equation (\ref{geodesic_like_point_curve}).
\item[The system of geodesic-like equations between two points:]

Moreover, if $\mathbf{c}_1$ and $\mathbf{c}_2$ are both constant curves on $S$, then the geodesic-like equations between these two points is
\begin{equation}\label{geodesic_like_two_points}
( \nabla E) = (
E_{u_1},E_{u_2},\cdots,E_{u_n-1},E_{v_1},E_{v_2},\cdots,E_{v_n-1} )
= 0.
\end{equation}

\end{description}

A curve satisfies one of equations (\ref{geodesic_like_two_curves})
- (\ref{geodesic_like_two_points}) is called a geodesic-like curve
between $\mathbf{c}_1$ and $\mathbf{c}_2$. Let us describe how to
find the local minimal geodesic-like curve between two curves
$\mathbf{c}_1$ and $\mathbf{c}_2$ on the surface $S$. In this
algorithm, we solve the system of geodesic-like curve equations by
the Newton's method and the iterator method.

\begin{Algorithm}\label{geodesic_like_algorithm}(Geodesic-like algorithm)
\begin{description}
\item[Step 1:]   Given two closed curves $\mathbf{c}_1$ and $\mathbf{c}_2$ on the surface. Input an initial curve $\mathbf{\alpha}$ such that the endpoints of $\mathbf{\alpha}$ are on
$\mathbf{c}_1$ and $\mathbf{c}_2$.

\item[Step 2:] Solving the geodesic-like equations (equation
(\ref{geodesic_like_two_curves}) or
(\ref{geodesic_like_two_points})) by the initial curve $\mathbf{\alpha}$ and
obtain a geodesic-like curve, which we still denote it by $\mathbf{\alpha}$,
between $c_1$ and $c_2$.

\item[Step 3:] If the set $(\mathbf{\alpha} \cap \mathbf{c}_1) \cup
( \mathbf{\alpha}\cap \mathbf{c}_2)$ consists of the  endpoints
of $\mathbf{\alpha}$, then $\mathbf{\alpha}$ is the local
minimal geodesic-like curve between $\mathbf{c}_1$ and
$\mathbf{c}_2$. Otherwise, trimming away some parts of the curve
$\mathbf{\alpha}$ such that the intersections of this trimmed
curve, which we still denote it by $\mathbf{\alpha}$, and
$(\bf{\alpha} \cap \mathbf{c}_1) \cup ( \mathbf{\alpha}\cap
\mathbf{c}_2)$ are only the endpoints of this trimmed curve (see
Figure \ref{figure_trimmed_curve}). Then repeat Step 2.
\end{description}
\end{Algorithm}

\begin{figure}
{\centering
\includegraphics[width=8cm]{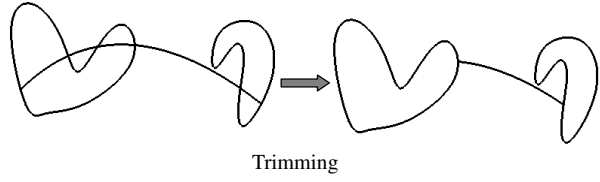}
\caption{the method of trimmed curves }\label{figure_trimmed_curve}
}
\end{figure}

By Theorem \ref{geodesic_like_approach_geodesic}, one will proceed by the geodesic-like algorithm to obtain the shortest path between $\mathbf{c}_1$ and $\mathbf{c}_2$ when the order $n$ is large enough. We summarize it as follows.

\begin{thm}\label{geodesic_like_approach_shortest}
Let $S$ be a parametric surface and $\mathbf{c}_1$, $\mathbf{c}_2$
be two closed curves on $S$. For each $n \geq 2$, $\tilde{\alpha}_n$
is the local minimal geodesic-like curve that obtained by the
geodesic-like Algorithm (algorithm \ref{geodesic_like_algorithm}).
If the set $\{ \tilde{\alpha}_n \}$ is a convergent sequence, then
there exists a local minimal geodesic $\gamma$ between
$\mathbf{c}_1$ and $\mathbf{c}_2$ such that
\begin{equation}
\lim_{n\rightarrow \infty} \tilde{\alpha}_n = \gamma.
\end{equation}
Moreover, $\tilde{\alpha}_n$ is orthogonal to $\mathbf{c}_1$ and $\mathbf{c}_2$ when $n$ is
large enough.
\end{thm}

\subsection{Periodic surfaces}

If $S$ is a periodic surface about one or two directions, then we
may not obtain the local minimal geodesic-like curve by the
Algorithm \ref{geodesic_like_algorithm}. It is because that the
minimality of geodesics may not be preserved by the map of
parametrization.  To avoid this problem, we shall rewrite the domain
$U$ of parametrization of $S$. For simplicity, we assume that the
original domain of parametrization is $U=[0,1] \times [0,1]$ and
then the parametrization $ \mathbf{x} :[0,1]\times[0,1] \rightarrow
S.$

\begin{description}
\item[One-directional periodic surface]
Assume that $S$ is a parametrization surface with u-directional
period and $\mathbf{x}$ is the parametrization on $S$. Its
domain of parametrization, the $uv$-plane, is as in Figure 2.
Then $\mathbf{x}(0,v) = \mathbf{x}(1,v)$ for each $v \in [0,1]$.
Hence there is a function $\tilde{\mathbf{x}} : \mathbb{R}
\times [0,1] \rightarrow S$ such that
$\tilde{\mathbf{x}}(\tilde{u},v) = \mathbf{x}(u,v)$ for some $u
\in [0,1]$ provided $(\tilde{u}-u)$ is an integer. Using the map
$\tilde{\mathbf{x}}$, we can find two curves $c_1^0$ and $c_1^1$
from $[0,1]$ to  $\mathbb{R}\times [0,1]$ such that
$\tilde{\mathbf{x}}(c_1^0([0,1])) =
\tilde{\mathbf{x}}(c_1^1([0,1])) =\mathbf{c}_1$ on $S$.
Moreover, we assume that $c_1^0 = c_1\subset [0,1]\times[0,1]$
and $c_1^1 \subset [1,2] \times[0,1]$. Using Algorithm
\ref{geodesic_like_algorithm}, the geodesic-likes curves from
$c_1^0$ to $c_2$and from $c_1^1$ to $c_2$ on $U$ can be found
and we denote them by $\gamma_0$ and $\gamma_1$ respectively
(see Figure \ref{figure_u_periodic}). Then the one in
$\{\mathbf{x}(\gamma_0),\mathbf{x}(\gamma_1)\}$ with smaller
length is the local minimal geodesic-like curve between
$\mathbf{c}_1$ and $\mathbf{c}_2$ on $S$.

\begin{figure}[h]
{\centering
\includegraphics[width=8cm]{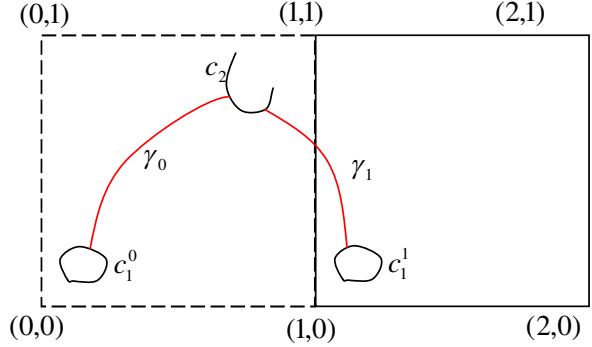}
\caption{Local minimal geodesic-like curve on a one-directional
periodic surface. }\label{figure_u_periodic} }
\end{figure}

\item[Two-directional periodic surface]
 If $S$ is a periodic surface about two directions, then we
 shall find  $\tilde{\mathbf{x}} : \mathbb{R}\times \mathbb{R}
 \rightarrow S$ such that
 $\tilde{\mathbf{x}}(\tilde{u},\tilde{v}) = \mathbf{x}(u,v)$ for
 some $u,v \in [0,1]$ if both $(\tilde{u}-u)$ and
 $(\tilde{v}-v)$ are integers. Similarly, we can find four
 curves $c_1^0 \subset [0,1] \times [0,1]$, $c_1^1 \subset [1,2]
 \times [0,1]$, $c_1^2 \subset [0,1] \times [1,2]$ and $c_1^3
 \subset [1,2] \times [1,2]$ (see Figure
 \ref{figure_uv_periodic}) such that $\mathbf{x}(c_1^i) =
 \mathbf{x}(c_1)=\mathbf{c}_1$ on $S$ for $i=0,1,2,3$. Denote
 the minimal geodesic-like curve between $c_1^i$ and $c_2$ by
 $\gamma_i$ for $i=0,1,2,3$. Thus the minimal geodesic-like
 curve on $S$ between $\mathbf{c}_1$ and $\mathbf{c}_2$ on $S$
 is the one in $\{\mathbf{x}(\gamma_i)\}_{i=0}^3$ with the
 shortest length.

 \begin{figure}[h]
{\centering
\includegraphics[width=8cm]{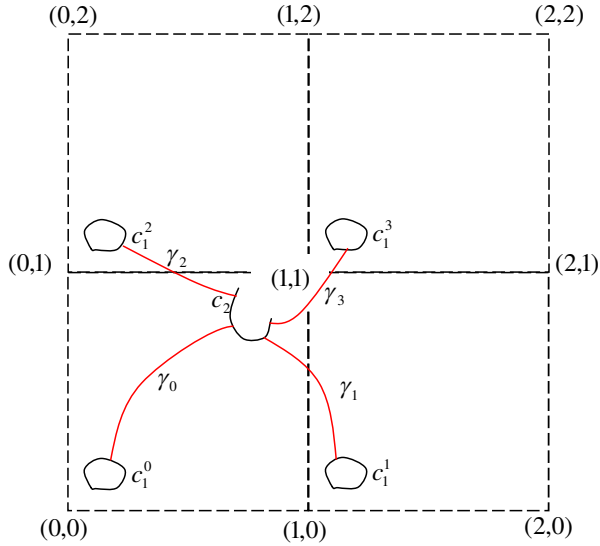}
\caption{Local minimal geodesic-like curve on a two-directional
periodic surface. }\label{figure_uv_periodic} }
\end{figure}
\end{description}

\section{Simulations}

To apply our method in practice, we present some examples by
simulation. The geodesic-like curves in our simulations are all
uniform quadratic B-spline curves in $\mathbb{R}^2$.

\begin{figure}
{\centering
\includegraphics[width=8cm]{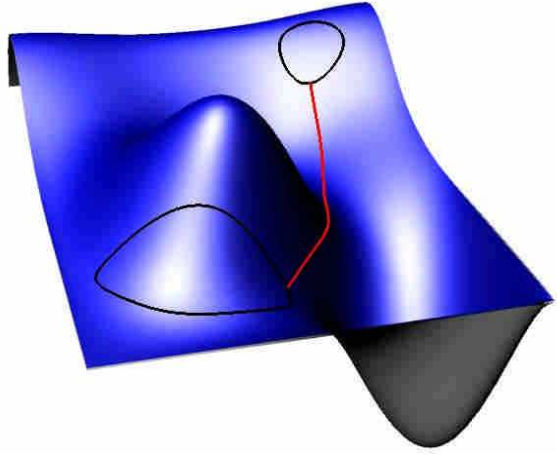}
\caption{Distance between two closed curves on a NURBS
surface}\label{figure_distance opensurface} }
\end{figure}

First we consider an open surface $S$ and two closed curves
$\mathbf{c}_1$ and $\mathbf{c}_2$ on $S$ as in Figure
\ref{figure_distance opensurface}. The surface $S$ is a cubic
B-spline surface with (8,4) control points. The red curve in figure
4 is the local minimal geodesic-like curve of order 11 between
$\mathbf{c}_1$ and $\mathbf{c}_2$ and  its error is less than
$10^{-6}$.

\begin{figure}[h]
{\centering
\includegraphics[width=8cm]{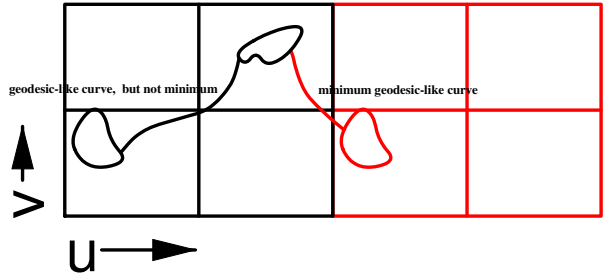}
\caption{domain of parametrization of a $u$-periodic periodic
surface}\label{figure_period_u} }
\end{figure}

\begin{figure}[h]
{\centering
\includegraphics[width=8cm]{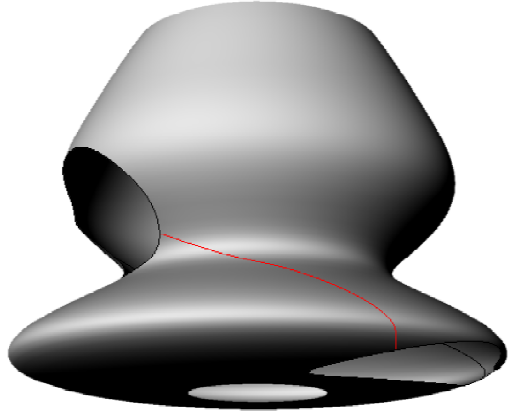}
\caption{One-directional periodic
surface}\label{figure_period_u_model} }
\end{figure}

Secondly, a surface of revolution is an example of one-dimensional
periodic surfaces. Figure \ref{figure_period_u} is the domain of
parametrization ($uv$-plane) of the surface of revolution as in
Figure \ref{figure_period_u_model}. In Figure \ref{figure_period_u},
there are two geodesic-like curves in the $uv$-plane, one is from
$c_1^0$ to $c_2$ and the other is from $c_1^1$ to $c_2$. Then the
image under parametrization of the shorter one is the local minimal
geodesic-like curve between these two curves on the surface.

\begin{figure}[h]
{\centering
\includegraphics[width=4cm]{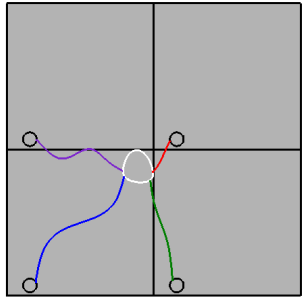}
\caption{Distance between two closed curves on torus
(uv-plane)}\label{figure_distance_torus_uvplane} }
\end{figure}

\begin{figure}[h]
{\centering
\includegraphics[width=9cm]{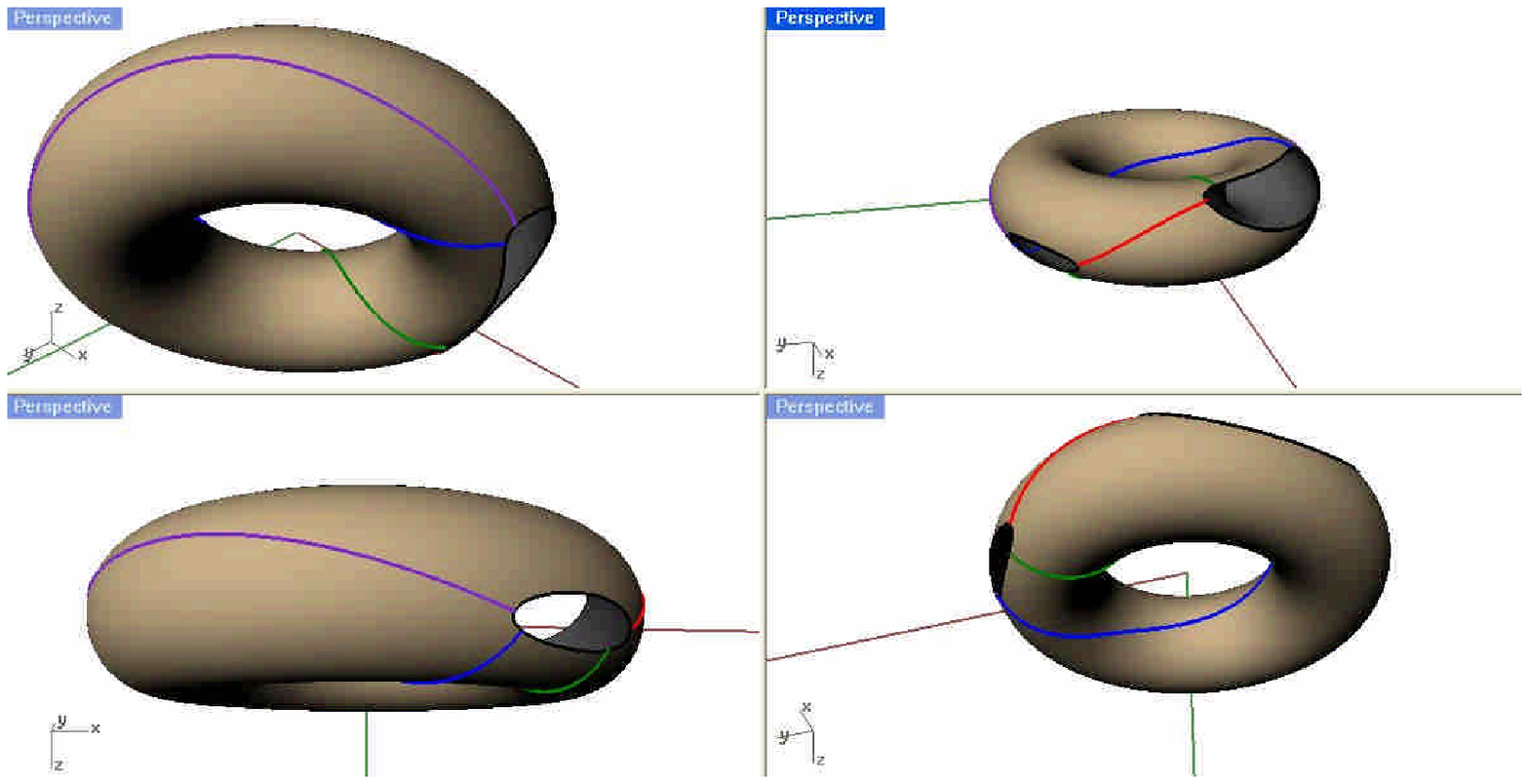}
\caption{Distance between two closed curves on
torus}\label{figure_distance_torus} }
\end{figure}

Thirdly, a typical example of two-dimensional periodic surfaces is
the torus. Figure \ref{figure_distance_torus_uvplane} is the domain
of parametrization of a torus as in Figure
\ref{figure_distance_torus}. There are four geodesic-like curves in
the $uv$-plane. Therefore, the image under parametrization of the
shortest one is the local minimal geodesic-like curve between two
closed curves on the torus.

\begin{figure}[h]
{\centering
\includegraphics[width=5cm]{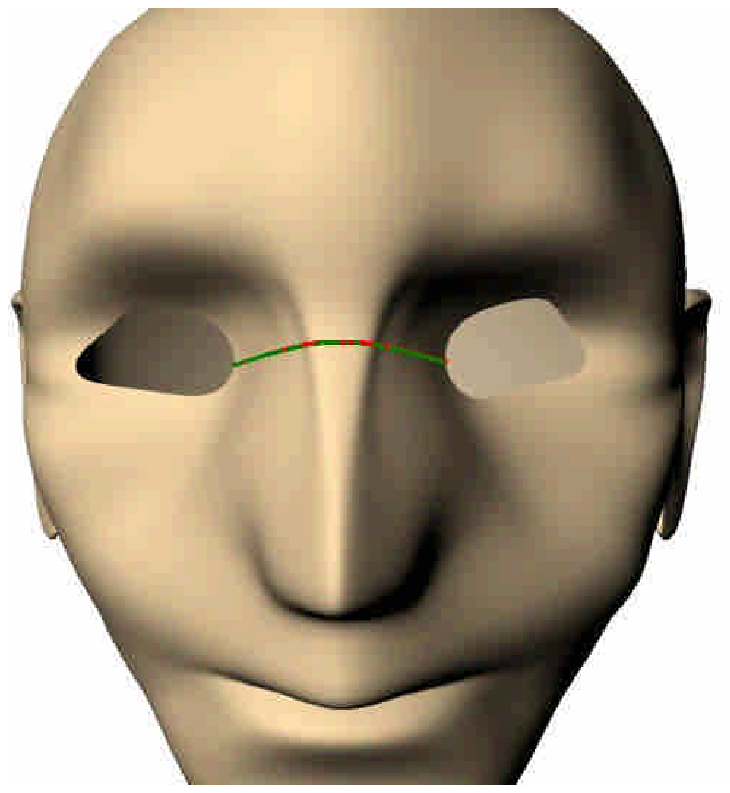}
\caption{The distance between two holds on a face model
}\label{figure_face_distance} }
\end{figure}

\begin{figure}[h]
{\centering
\includegraphics[width=13cm]{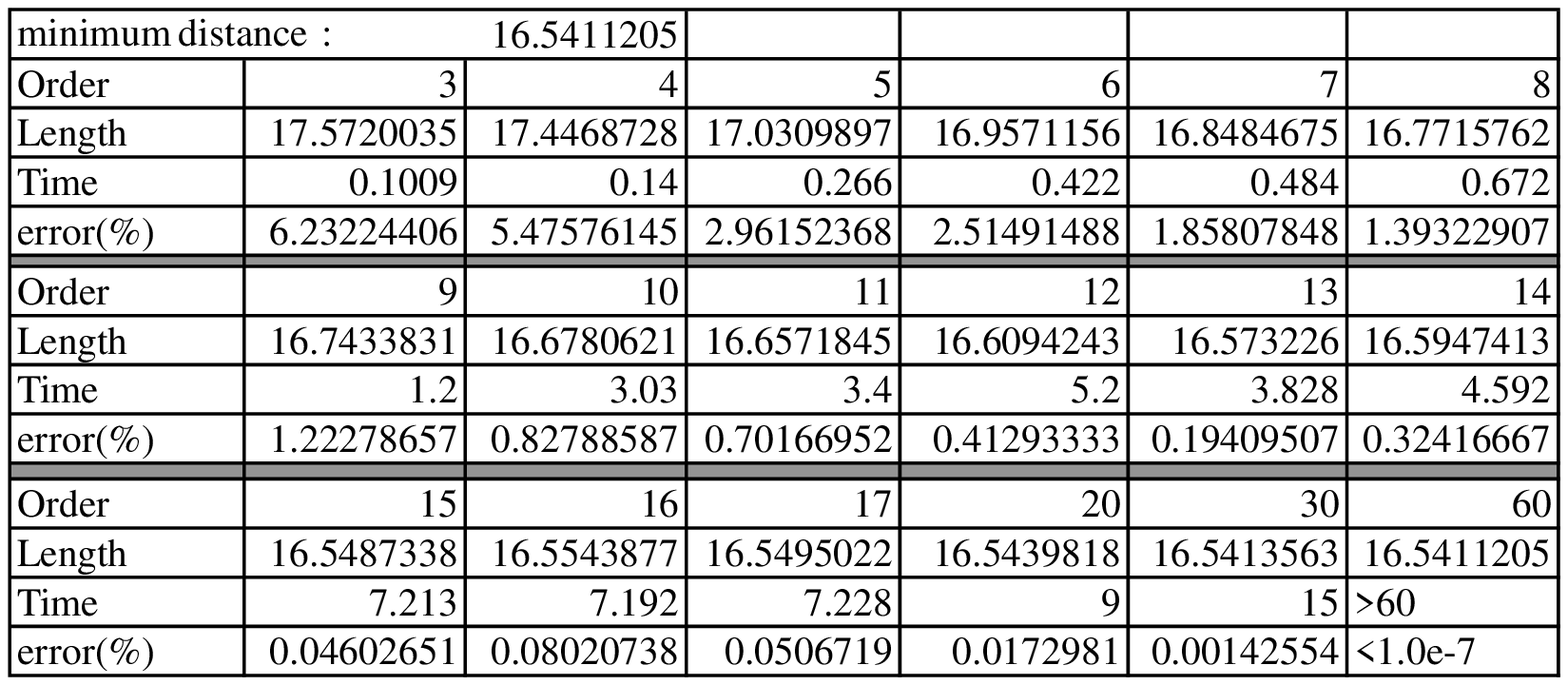}
\caption{The table of the distance between two holds on a face model
with different orders }\label{figure_table_face_distance} }
\end{figure}

Lastly, we construct a face model as in Figure
\ref{figure_face_distance} by NURBS surface and find the minimal
geodesic-like curves between two holds (the eyes) on the surface.
The data in Figure\ref{figure_table_face_distance} are about the
geodesic-like curves of different orders between the two holes in
Figure 9. Here in Figure \ref{figure_table_face_distance} the order
means the number of control points while $\mbox{error(\%) }$ is the
percentage of error, which is defined by
\begin{equation}
\mbox{error(\%) }=\frac{\mbox{ Length }- \mbox{ minimum
distance}}{\mbox{ minimum distance}} \times 100\%.
\end{equation}
The red curve in Figure \ref{figure_face_distance} is the local
minimal geodesic-like curve of order 30 and the green curve is the
exact minimal geodesic between two holes. Then the lengths of
geodesic-like curves constructed by our method approaches the
minimum distance between the two holes. To deserve to be mentioned,
$\mbox{error(\%) }$ will be less than $10^{-7}$ provided the
geodesic-like curve is constructed by 60 control points. It proposes
that the geodesic-like algorithm has increased actually
computational efficiency of this simulation .

\section{Discussion}
The geodesic-like algorithm provides an effective and reliable
computation of shortest paths between two curves on surfaces. For
computing the shortest paths between two curves on $\mathbb{R}^3$,
our method is comparable with other well-known methods. Especially,
the construction of geodesic-like curves only bases on the uniform
quadratic B-spline curves since it is enough to us to consider the
geodesic-like curves in the plane. Significatively, our method can
be extended to solve the distance problem between any two objects on
surfaces and the distance problem in higher dimension.

To solve the system of geodesic-like equations, however, Newton's
method is too expansive. Moreover, it can only solve local minimal
geodesic-like curves but not global minimum ones. In the future
investigation, we expect to find a numerical method to solve
efficiently all local minimal geodesic-like curves between two
objects on surfaces to overcome these problems.

\end{document}